\documentclass[a4paper,UKenglish,cleveref, autoref, thm-restate, numberwithinsect, pdfa]{lipics-v2021}

\pdfoutput=1
\usepackage{booktabs}
\usepackage{graphicx}
\usepackage{xcolor}
\usepackage{tikz}
\usetikzlibrary{backgrounds}
\usepackage{float}
\usepackage{xcolor}
\usepackage[framemethod=tikz]{mdframed}
\usepackage{array}
\usepackage{ragged2e}
\usepackage{comment}

\usepackage{tabularx}
\newcolumntype{Y}{>{\RaggedRight\arraybackslash}X}

\newcommand{\repo}[1]{\parbox[t]{2.3cm}{\raggedright\ttfamily #1}}

\usepackage{longtable}
\usepackage{rotating}
\usepackage{pgfplots}
\pgfplotsset{compat=1.18}

\definecolor{background-gray}{gray}{0.96}

\newmdenv[
  topline=false,
  bottomline=false,
  rightline=false,
  leftline=true,
  linecolor=black,
  linewidth=3pt,
  backgroundcolor=background-gray,
  skipabove=6pt,
  skipbelow=6pt,
  innertopmargin=6pt,
  innerbottommargin=6pt,
  innerleftmargin=10pt,
  innerrightmargin=10pt
]{results}

\title{An Exploratory Study of Agent Plans for Agentic AI Coding Tools in Open-Source Software}

\titlerunning{Agent Plans in Open-Source Software}

\author{Muhammad Auwal Abubakar}{
University of Bamberg, Germany
}{muhammad.abubakar@uni-bamberg.de}{0009-0006-1028-0650}{}

\author{Seyedmoein Mohsenimofidi}{
Heidelberg University, Germany
}{s.mohsenimofidi@uni-heidelberg.de}{0009-0009-1620-2735}{}

\author{Jai Lal Lulla}{
Singapore Management University, Singapore
}{jailal.l.2025@phdcs.smu.edu.sg}{0009-0005-0024-8238}{}

\author{Jie M. Zhang}{
King’s College London, United Kingdom
}{jie.zhang@kcl.ac.uk}{0000-0003-0481-7264}{}

\author{Christoph Treude}{
Singapore Management University, Singapore
}{ctreude@smu.edu.sg}{0000-0002-6919-2149}{}

\author{Sebastian Baltes}{
Heidelberg University, Germany
}{sebastian.baltes@uni-heidelberg.de}{0000-0002-2442-7522}{}

\author{Matthias Galster} {
University of Bamberg, Germany
}{mgalster@ieee.org}{0000-0003-3491-1833}{}

\authorrunning{M.\,A. Abubakar et al.} 

\Copyright{Muhammad Auwal Abubakar, Seyedmoein Mohsenimofidi, Jai Lal Lulla, Jie M. Zhang, Christoph Treude, Sebastian Baltes, and Matthias Galster} 

\ccsdesc{Software and its engineering~Software creation and management}
\ccsdesc{Software and its engineering~Software configuration management and version control systems}
\ccsdesc{Computing methodologies~Artificial intelligence} 

\keywords{Agentic AI Coding Tools, Configuration Mechanisms, Plans}

\category{Emerging Results, Vision \& Reflection Track Paper}

\relatedversion{} 

\nolinenumbers 

\EventEditors{Robert Feldt, Maria Paasivaara, Daniel Mendez, Stefan Wagner, and Marvin Mu\~{n}oz Bar\'{o}n}
\EventNoEds{5}
\EventLongTitle{20th International Symposium on Empirical Software Engineering and Measurement (ESEM 2026)}
\EventShortTitle{ESEM 2026}
\EventAcronym{ESEM}
\EventYear{2026}
\EventDate{October 8--9, 2026}
\EventLocation{Munich, Germany}
\EventLogo{}
\SeriesVolume{394}
\ArticleNo{53}

\newcommand{\matthias}[1]{\textcolor{magenta}{{\it [Matthias says: #1]}}}

\begin{document}
\maketitle

\begin{abstract}
Repository-level configuration artifacts allow developers to provide guidance for agentic AI coding tools, e.g., Claude Code, Gemini. Although prior research has examined repository-shared context files that capture project-level instructions and conventions (e.g., AGENTS.md files), little is known about more task-oriented artifacts such as \emph{Agent Plans}. 
We present an exploratory study of Agent Plans in open-source software repositories, examining how plan files are preserved, which development activities they support, and what information they provide to guide agent execution. We screened 36{,}710 GitHub repositories belonging to engineered software projects and identified 85 Markdown plan files from 10 repositories. 
Within this concentrated corpus, Agent Plans supported several kinds of software engineering work, including maintenance, design, construction, 
quality-related work, 
and process support. They also provided task-oriented execution guidance, most commonly through implementation steps, concrete files and locations, and testing and validation information. Overall, repository-preserved Agent Plans in these tool-specific directories appear to be a narrow but informative artifact for studying task intent and execution guidance in human-agent workflows.
\end{abstract}

\section{Introduction}
\label{sec:typesetting-introduction}

Repository-level configuration artifacts allow developers to guide agentic AI coding tools, such as Claude Code ~\cite{galster2026configuring,mohsenimofidi2025context}. Prior work has shown that artifacts such as \texttt{AGENTS.md} files capture project knowledge, development conventions, and workflow-related guidance that is versioned and inspectable within repositories~\cite{galster2026configuring,mohsenimofidi2025context,chatlatanagulchai2025agent}. This means that interactions with agentic AI coding tools are shaped not only by prompts, but also by persistent repository-level artifacts. 
Prior work has identified \emph{configuration mechanisms} for such tools, that is, general ways of tailoring tool or agent behavior to a project or workflow; examples include Context Files, Skills, Subagents, etc.~\cite{galster2026configuring}. \emph{Context files} are one particularly prominent type of configuration mechanism whose artifacts are Markdown files, such as \texttt{AGENTS.md}, \texttt{CLAUDE.md}, or \texttt{copilot-instructions.md}. These files provide persistent project-level instructions, conventions, and contextual guidance for agentic AI coding tools. Such guidance may include build and run commands, test procedures, architectural information, coding conventions, and project-specific workflow rules~\cite{galster2026configuring,chatlatanagulchai2025agent}. Thus, interactions with agentic AI coding tools are shaped not only by prompts, but also by persistent repository-level guidance. Earlier research has largely examined context files at a \emph{coarse-grained level}, asking whether such files  exist, how they are adopted, and whether they improve tool performance~\cite{galster2026configuring,chatlatanagulchai2025agent,mohsenimofidi2025context,gloaguen2026evaluating,lulla2026impact}. 
However, \emph{task-oriented} planning artifacts remain underexplored.

In this study, \emph{Agent Plans} refer to task-oriented Markdown artifacts that describe a concrete unit of intended work for an agentic AI coding tool. Plan files are used with tools such as Cursor, where the agent can inspect the codebase, identify relevant files, and produce a plan file for the developer to review, edit, or approve the plan files before implementation~\cite{cursorPlanMode}. Agent Plans are therefore hybrid human-agent artifacts: they may be drafted by the agent, reviewed or revised by the developer, and then used to guide execution. When saved as Markdown files in a repository, they become accessible to other developers or future agents. For example, the \textit{travis-php85-gha-migration} plan in \texttt{boldgrid/w3-total-cache}~\cite{w3cacheMigrationPlan} describes a staged migration from Travis CI to GitHub Actions and later Argo Workflows, identifies migration-related files such as \texttt{.travis.yml}, \texttt{composer.json}, \texttt{phpunit.xml}, and planned GitHub Actions workflows, and specifies validation work such as running PHPUnit and proving the replacement CI workflow before retiring the Travis CI workflow. Unlike general-purpose context files such as \texttt{AGENTS.md}, which typically provide persistent project knowledge or behavioral instructions, Agent Plans are explicitly oriented toward concrete development tasks and ongoing implementation activity. We therefore treat Agent Plans as a distinct form of repository-based configuration artifact. Importantly, not all Agent Plans are preserved in public repository history; some may remain local artifacts on a developer’s machine. To characterize how Agent Plans manifest as repository artifacts and what role they play in developer-agent collaboration, this study investigates three research questions in the context of open-source software (OSS) systems:

\begin{description}
\item[RQ1:] \textbf{How are Agent Plans preserved as repository artifacts in OSS?}
\item[RQ2:] \textbf{What tasks and developer activities do Agent Plans support?}
\item[RQ3:] \textbf{What information do Agent Plans provide to guide task execution?}
\end{description}

These questions are important because Agent Plans sit between a developer's task request and the agent's execution. In the planning workflow, the agent may draft a plan file in response to the developer's request, the developer can then read, revise, or approve it before the agent proceeds with the work. When plan files are saved in a repository, they become visible records of human-agent collaboration rather than remaining local or transient interaction artifacts. By focusing on these repository-preserved Agent Plans, this paper examines how development work is organized for agentic AI coding tools beyond broader repository-level context artifacts. 
When committed, they provide observable evidence of how a development task was framed, structured, and prepared for execution by agents. We focus on this directly inspectable subset of Agent Plans while excluding plans that remained local or were stored outside the searched directories.
This paper makes three contributions:
\begin{itemize}
    \item First, it provides an exploratory empirical study of Agent Plans as repository-preserved artifacts in OSS development. 
    \item Second, it shows that Agent Plans in the observed corpus support activities including maintenance, design, construction, quality-related work, and process support; requirements clarification and design exploration were not observed in this corpus. 
    \item Third, it shows that Agent Plans mostly specify implementation steps, files and locations, and testing and validation information. 
\end{itemize}
Together, these contributions help clarify what Agent Plans are as observable repository artifacts, what kinds of work they capture, and how they differ from project-level context files. The remainder of this paper is organized as follows. Section~\ref{sec:background} reviews background and related work on agentic AI coding tools and repository-level context artifacts. Section~\ref{sec:data_collection_analysis} describes the data collection and analysis procedures. Section~\ref{sec:results} presents the results for the three research questions. Section~\ref{sec:discussion} discusses the implications of the study and its threats to validity. Finally, Section~\ref{sec:conclusion} concludes the paper.

\section{Background and Related Work}
\label{sec:background}


Studies of GitHub repositories and pull requests report a growing use of agentic AI coding tools and agent-generated contributions across activities such as bug fixing, feature implementation, and documentation updates~\cite{robbes2026agentic,li2026aidev,watanabe2025use}. Related work also shows that LLM-based coding tools can favor popular technology choices even when they are not well suited to the task~\cite{twist2026preferences}, highlighting the need for project-specific guidance. A second line of research examines how developers incorporate coding agents into collaborative practice. Cynthia et al.~\cite{cynthia2026we} show that agent use differs between developer roles: peripheral contributors rely more heavily on agents for implementation, whereas core developers use them more selectively and focus on review and validation. This shifts attention from whether agents are adopted to how agentic coding is coordinated. A third line of research studies repository-level artifacts that guide agentic AI coding tools. Galster et al.~\cite{galster2026configuring} identify mechanisms for configuring such tools, including prompts, instruction files, and repository-level context files. Related studies examine context files and agent-facing repository documentation in more detail, including \texttt{AGENTS.md} files and other files that encode conventions, architecture, build commands, test execution, setup, and workflow guidance~\cite{chatlatanagulchai2025agent,mohsenimofidi2025context}. Experimental studies report mixed evidence about their effects: context files can improve agent performance in some settings, but their impact depends on the model, task, and evaluation setting~\cite{gloaguen2026evaluating,lulla2026impact}.
 
This body of work shows that agentic AI coding tools are becoming part of OSS development and that repository-level files can influence how these tools operate. Also, prior work has mainly focused on broad project-level context files or agent-generated code contributions. Less is known about task-oriented plan files that describe concrete units of intended work. Agent Plans capture how a task is framed before or during execution, including the intended changes, relevant files or locations, and checks used to assess completion. This study addresses this gap. Agent Plans follow a hybrid human-agent workflow. 
Support for this workflow has emerged in major tools including Cursor, Gemini, and Claude Code. When developers commit plan files to a repository, typically under directories such as \texttt{.cursor/plans} or \texttt{.claude/plans}, they become inspectable by other developers and agents, functioning as a lightweight record of intended work alongside the resulting code.

\section{Data Collection and Analysis}
\label{sec:data_collection_analysis}

\subsection{Data Collection}
\label{sec:data_collection}

\begin{figure}[H]
    \centering
    \includegraphics[width=1\linewidth]{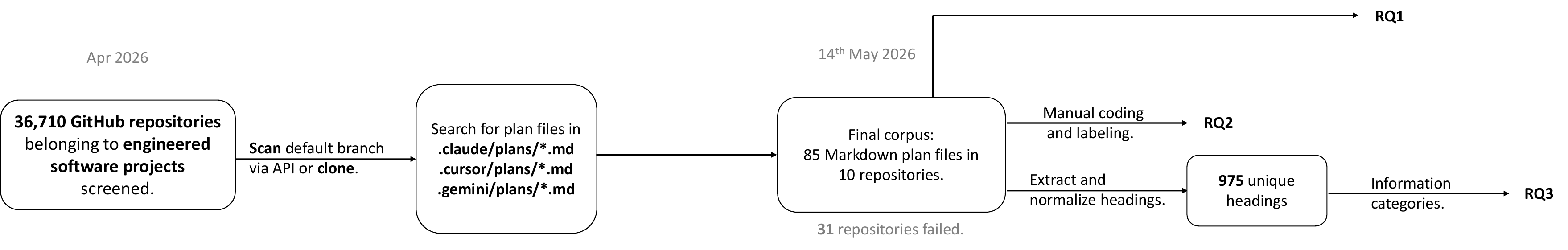}
    \caption{\textbf{Process for identifying Agent Plan Markdown files in GitHub repositories.}}
    \label{fig:data-collection}
\end{figure}

Figure~\ref{fig:data-collection} summarizes the data collection process. The repository scan was conducted on May 14, 2026. The results therefore describe the plan files present on the repository's default branches at that point in time. The set of repositories screened in this study was derived from the repository processing pipeline introduced by Baltes et al.~\cite{galster2026datasetpaper}. The pipeline began with a metadata-filtered set of 40{,}585 GitHub repositories and subsequently classified 36{,}710 of them as belonging to engineered software projects (i.e., repositories of sufficient size, age, activity, and with indicators that point to the use of software engineering practices)~\cite{DBLP:journals/ese/MunaiahKCN17}. 

To identify Agent Plan artifacts, we scanned the default branches of these repositories using the GitHub API. 
The search targeted Markdown files stored under tool-specific plan directories: \texttt{.claude/plans/*.md}, \texttt{.cursor/plans/*.md}, and \texttt{.gemini/plans/*.md}. A repository could contain more than one such file, and each individual Markdown file was treated as a separate unit of collection and analysis. The documentation for Cursor and Claude Code establishes the existence of planning workflows, but does not provide a complete convention for storing plan files in the repository~\cite {cursorPlanning,claudePlanMode}. We therefore used the observed tool-specific plan-directory patterns as conservative search locations for repository-preserved planning artifacts, while recognizing that plan files stored under other paths would not be captured. This directory-based strategy increases precision by focusing on explicit tool-plan locations, but it can miss plan-like artifacts stored under general paths such as \texttt{docs/}, \texttt{tasks/}, \texttt{specs/}, or issue and pull-request bodies. The \texttt{.gemini/plans/*.md} pattern was included to keep the search strategy symmetric across the agentic coding tools considered in the study. It returned no files among the 36{,}710 repositories screened. We therefore treat this as a null result for the search, not as evidence that Gemini-based planning artifacts are absent or that Gemini users do not create plans. 
Repositories without matching plan files were excluded. 

During processing, 31 repositories could not be retrieved or processed successfully and were recorded as failed repositories. For repositories with candidate plan files, the files were retrieved along with repository- and file-level metadata. 
After retrieval, candidate files were manually inspected to confirm that they were, in fact, plan files, that is, task-oriented Markdown documents describing the intended development work for an agentic AI coding tool. This validation step was a file-inclusion check rather than the qualitative coding used for RQ2 and RQ3. Because all retrieved files came from highly specific tool-plan directories and all met the inclusion criterion, no candidate files were removed at this stage. 
All candidate files met this criterion, which was expected given the specificity of the tool-specific directory patterns used in the search. Thus, because the search was restricted to predefined tool-specific directory conventions, plan-like artifacts stored under other names or locations may have been missed. After validation, the collected files were assembled into a local corpus while preserving their repository context.

The final corpus consisted of 85 Markdown plan files drawn from 10 repositories. These files formed the empirical basis for the subsequent analysis. 
Individual plan files ranged from 14 to 901 lines ($M = 178$) and 116 to 5{,}227 words ($M = 876$), with 1 to 41 section headings to structure plan files ($M = 13.5$), totaling 1{,}145 heading occurrences. 

\subsection{Data Analysis}
\label{sec:data_analysis}

To address \textbf{RQ1}, we examined how Agent Plans were preserved in the repository history. We counted repositories that contain plans, plan files per repository, and associated repository- and file-level metadata, including commit counts and contributor summaries.

To address \textbf{RQ2}, we used iterative qualitative coding to identify the tasks and developer activities reflected in the 85 plan files~\cite{schreier2012qca,mayring2014qca}. The coding unit was one complete plan file. For each file, coders first inspected the file name, title, and opening material, such as summaries, goals, or problem statements, where present. They then checked this interpretation against the remaining content, including implementation notes, file lists, verification steps, and to-do items. This procedure was used to avoid assigning labels based only on a single heading or isolated section. The coding framework had three levels. The \texttt{Task}-level captured the concrete task described in the plan file. The \texttt{Task-group}-level grouped related tasks into recurring activities, such as refactoring or testing. The \texttt{SWEBOK}-level mapped task groups to SWEBOK knowledge areas as higher-level representations of software engineering practices and technical, operational, and managerial activities~\cite{bourque2014swebok}. The final coding was bottom-up: each plan file was assigned a \texttt{Task}-level label, then a \texttt{Task-group}-level label, and finally one \texttt{SWEBOK} category. When a plan file contained multiple possible activities, coders assigned the label that best captured the plan's primary task based on the plan title, stated goal, and dominant implementation content. We used a primary-task label rather than multi-label coding to keep each plan file as a comparable unit in the SWEBOK and task-group distributions; this reduces double-counting but may underrepresent secondary activities. We therefore interpret the RQ2 results as distributions of primary plan purposes rather than exhaustive lists of every activity mentioned in the files.

The final coding involved two researchers. One researcher developed the initial coding framework and coded the full corpus; a second researcher independently coded the same 85 plan files at the \texttt{Task}, \texttt{Task-group}, and \texttt{SWEBOK} levels using the coding guide. The two coding sets were matched by plan file link, and all 85 plans were present in both sets. For \texttt{SWEBOK}-level labels, the coders achieved Cohen's \(\kappa = 0.7586\) with 80.0\% raw agreement. For \texttt{Task-group}-level labels, after applying a predefined mapping between equivalent label names, they achieved Cohen's \(\kappa = 0.8002\) with 82.4\% raw agreement. Both correspond to substantial agreement~\cite{landis1977measurement}. Disagreements were discussed and resolved to produce the final coding used in the results.

To address \textbf{RQ3}, we extracted all Markdown section headings from the 85 plan files, normalized them by lowercasing, removing special characters, lemmatizing, and conservatively merging lexical variants with the same meaning. All 85 plan files included headings. In total, we observed 1{,}145 heading occurrences and 975 distinct representative headings. Following related work on \texttt{AGENTS.md} and \texttt{README} files~\cite{mohsenimofidi2025context,gao2025adapting}, we limited the information-category analysis to level-1 and level-2 headings, which comprised 592 occurrences: 126 at level 1 and 466 at level 2.
The coding unit was a level-1 or level-2 section, consisting of the heading and its section content, including prose, bullet points, file paths, commands, code snippets, tables, and nested lower-level subsections where present. Headings were used to locate and group recurring sections, but information categories were assigned by reading the corresponding section content. We developed an initial coding guide from six representative headings that appeared in at least three repositories and at least three times at levels 1 or 2, then applied it to 23 representative headings that appeared in at least two repositories and at least twice. 
The resulting coding guide covered nine information categories, which are reported in Table~\ref{tab:plan-information-categories} and discussed in Section~\ref{sec:results_rq3}. One researcher developed the initial category mapping, and a second researcher independently coded the 472 representative level-1 and level-2 headings using the coding guide. Equivalent category names were mapped before calculating agreement. The heading-level coding achieved Cohen's \(\kappa = 0.7359\) with 77.1\% raw agreement, corresponding to substantial agreement~\cite{landis1977measurement}. Ambiguous cases and category names were discussed and resolved by returning to the corresponding section content.

\section{Results}
\label{sec:results}

\subsection{RQ1: How are Agent Plans Preserved as Repository Artifacts?}
\label{sec:results_rq1}


Agent Plans were rarely found under canonical tool-specific directories in public repository history. This is consistent with plan preservation being an emerging practice: Cursor introduced Plan Mode only in early October 2025. Among 36{,}710 GitHub repositories, we found 85 plan Markdown files across ten repositories (Table~\ref{tab:rq1-repo-distribution}), stored under \texttt{.cursor/plans} (\(n = 72\)) and \texttt{.claude/plans} (\(n = 13\)). One possible reason for committing such files is to make implementation decisions visible to collaborators and code reviewers, preserving task reasoning alongside the code.


\begin{table}[H]
\centering
\caption{Distribution of Agent Plan files with selected repository-level characteristics.}
\label{tab:rq1-repo-distribution}
\setlength{\tabcolsep}{2.5pt}
\begin{tabularx}{\textwidth}{>{\RaggedRight\arraybackslash}Xrrrrr}
\toprule
\textbf{Repository} & \textbf{\#Plan files} & \textbf{Created} & \textbf{\#Contrib.} & \textbf{\#Commits} & \textbf{Size (KB)} \\
\midrule
\texttt{forcedotcom/salesforcedx-vscode} & 66 & 2017 & 119 & 5,013 & 906,966 \\
\texttt{youtrackdb/youtrackdb} & 8 & 2024 & 143 & 24,278 & 262,502  \\
\texttt{forcedotcom/sfdx-core} & 3 & 2018 & 57 & 3,553 & 79,282  \\
\texttt{adobe/spectrum-design-data} & 2 & 2022 & 14 & 1,337 & 46,196 \\
\texttt{hubtype/botonic} & 1 & 2018 & 34 & 3,297 & 49,664  \\
\texttt{nam20485/OdbDesign} & 1 & 2023 & 6 & 2,230 & 10,459   \\
\texttt{natemcmaster/commandlineutils} & 1 & 2017 & 99 & 1,030 & 18,073  \\
\texttt{nomadkaraoke/python-audio-separator} & 1 & 2023 & 26 & 353 & 150,754  \\
\texttt{boldgrid/w3-total-cache} & 1 & 2018 & 49 & 1,203 & 23,632 \\
\texttt{input-output-hk/daedalus} & 1 & 2016 & 62 & 11,099 & 255,858 \\
\bottomrule
\end{tabularx}
\end{table}

The corpus was strongly concentrated. \texttt{forcedotcom/salesforcedx-vscode} accounted for 66 of the 85 plan files. We report sensitivity checks for RQ2 and RQ3, excluding this repository. Plan-file preservation was also contributor-concentrated: \texttt{mshanemc} contributed 64 of the 66 plan files in \texttt{forcedotcom/salesforcedx-vscode} and all three in \texttt{forcedotcom/sfdx-core}; \texttt{andrii0lomakin} contributed all eight in \texttt{youtrackdb/youtrackdb}; and \texttt{GarthDB} contributed both files in \texttt{adobe/\allowbreak spectrum-design-data}. 

The remaining repositories likewise had a single contributor for preserved plan files, except \texttt{input-output-hk/\allowbreak daedalus}, where the single plan file involved two contributors. However, committing Agent Plans was not a routine practice among these contributors across their repositories. Instead, contributors who committed plan files in one repository did not necessarily do so in others. To distinguish intentional sharing from incidental preservation, we checked whether each plan file was revised, referenced in repository issues or pull requests, or followed by adding the plan directory to \texttt{.gitignore}. Most plan files (\(n=78\)) appeared in one single commit, and 73 of these had no observed issue or pull-request reference. Only seven files were revised across multiple commits, and only two of those were also externally referenced. No repository later added the plan-file directory to \texttt{.gitignore}. Overall, most committed plan files are therefore better characterized as ephemeral task records than as intentionally maintained project artifacts.


To assess AI tool co-authorship, we examined 72 unique commits that touched plan files and for which complete GitHub metadata were available. Of these, 31 commits contained explicit AI co-authorship signals, affecting 27 of the 85 plan files. We counted only explicit \texttt{Co-authored-by} lines naming Claude or Cursor; tool-name mentions in commit messages were not treated as authorship evidence. Thus, we distinguish plan files associated with explicit AI-co-authored commits from those with no explicit AI co-authorship signal, without treating the latter as confirmed human-authored. These signals are conservative commit-level evidence: repository metadata does not show whether a plan was drafted by an agent, edited by a developer, used during implementation, or committed incidentally.


Revision behavior was similar across the two groups. Only 2 of the 27 AI-co-authored plan files and 5 of the 58 files without explicit AI co-authorship signals were revised after their first plan-file commit. Issue or pull-request references appeared in plan-file commit messages for 23 of the 27 AI-co-authored files and 54 of the 58 files without explicit AI co-authorship signals. Thus, AI co-authorship affects how the corpus should be interpreted, but it did not correspond to a distinct revision pattern.

\subsection{RQ2: What Developer Activities do Agent Plan Files Support?}
\label{sec:results_rq2}

\begin{figure}[htbp]

    \centering
    \includegraphics[width=\linewidth]{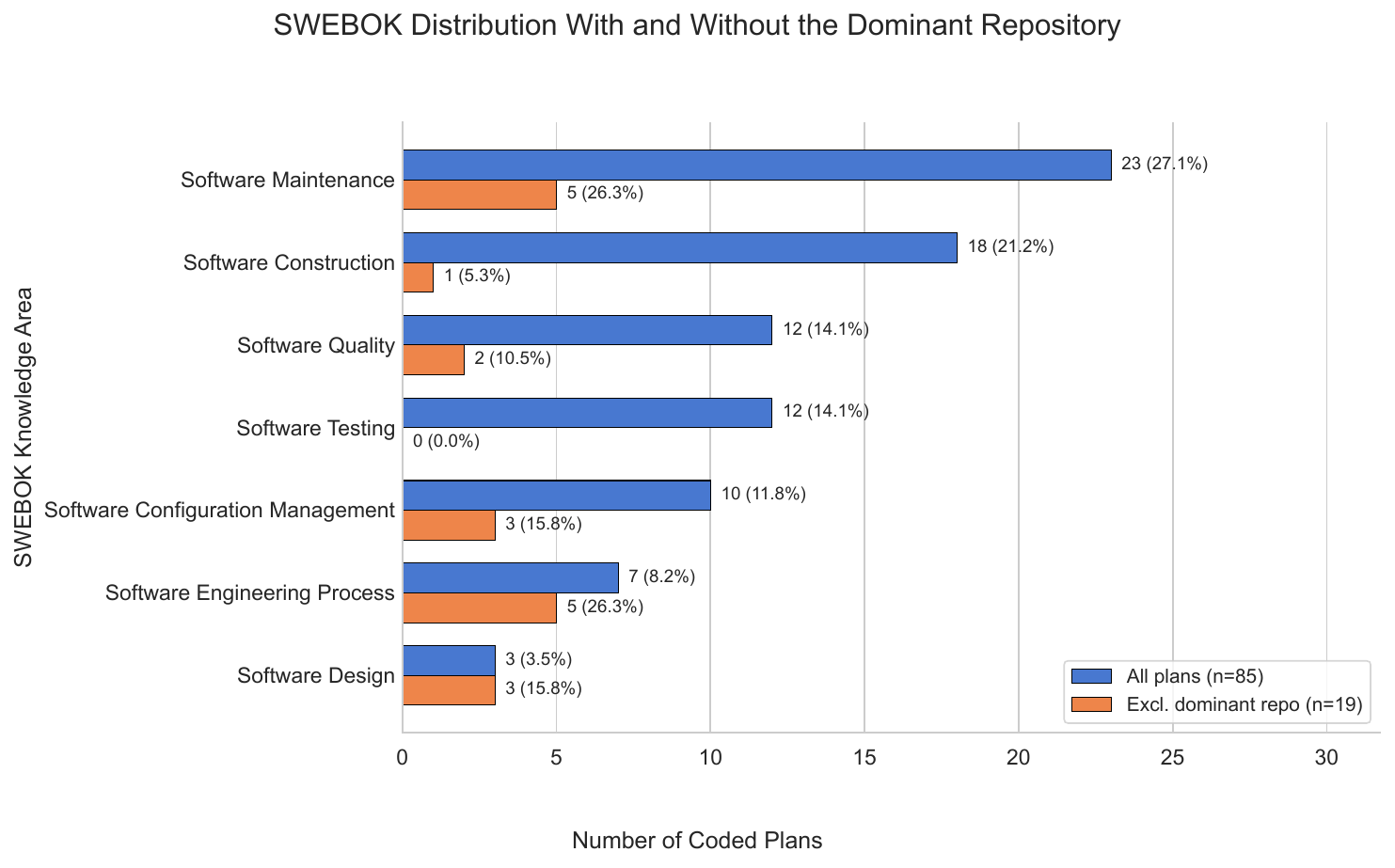}
    \caption{Distribution of Agent Plans across SWEBOK knowledge areas. 
    }
    \label{fig:rq2-swebok-distribution}
\end{figure}

Using the three-level coding framework described in Section~\ref{sec:data_analysis}, we coded each of the 85 Agent Plan files with one concrete \texttt{Task}-level label, one \texttt{Task-group} level label, and one \texttt{SWEBOK} knowledge area. Figure~\ref{fig:rq2-swebok-distribution} shows the SWEBOK distribution for the full corpus and, as a sensitivity check, the distribution after excluding the dominant repository, \texttt{forcedotcom/salesforcedx-vscode}.

In the full corpus, the most frequent SWEBOK areas were \emph{Software Maintenance} (\(n=23\)) and \emph{Software Construction} (\(n=18\)), followed by \emph{Software Testing} and \emph{Software Quality} (\(n=12\) each). Thus, the observed plan files were not limited to new-feature implementation; they also supported refactoring, migration, bug fixing, test creation, test repair, and quality enforcement. Figure~\ref{fig:rq2-swebok-distribution} also reports the distribution after excluding the dominant \texttt{forcedotcom/salesforcedx-vscode} repository. The remaining 19 plans still covered several SWEBOK areas, but the distribution changed substantially: \emph{Software Testing} was not observed among the 19 plan files outside the dominant repository, \emph{Software Construction} dropped to one plan file, and \emph{Software Engineering Process} and \emph{Software Design} became more prominent relative to the smaller remaining corpus. Therefore, the aggregate RQ2 findings should be interpreted as describing the observed corpus rather than as population-level evidence about all Agent Plans in open-source repositories. Because only 19 plan files remain after excluding the dominant repository, the sensitivity analysis should be read as a robustness check to assess whether categories still appear outside that repository, rather than as a representative cross-repository estimate.

We also split the RQ2 labels by explicit AI co-authorship signal, distinguishing plan files associated with commits containing such signals from those without. Both groups covered multiple SWEBOK areas. Plan files associated with AI-co-authored commits were most frequent in \emph{Software Maintenance} and \emph{Software Construction} (\(n=7\) each), followed by \emph{Software Testing} and \emph{Software Configuration Management} (\(n=4\) each), \emph{Software Design} (\(n=3\)), \emph{Software Quality} and \emph{Software Engineering Process} (\(n=1\) each). Plan files without explicit AI co-authorship signals were also concentrated in \emph{Software Maintenance} (\(n=16\)) and \emph{Software Construction} (\(n=11\)), followed by \emph{Software Quality} (\(n=11\)) and \emph{Software Testing} (\(n=8\)), \emph{Software Configuration Management} and \emph{Software Engineering Process} (\(n=6\) each). Thus, this split does not suggest that AI-co-authored plan files supported a separate type of software engineering work.

At the task-group level, the most frequent category was \emph{refactoring and migration} (\(n=20\), after harmonizing capitalization and equivalent labels), followed by \emph{testing and validation} (\(n=12\)), \emph{feature development} (\(n=12\)), \emph{build and CI/CD workflow} (\(n=10\)), \emph{code quality} (\(n=8\)), \emph{developer tooling} (\(n=7\)), and \emph{bug fixing} (\(n=6\)). Outside the dominant repository, \emph{testing and validation} was not observed, while \emph{developer tooling} (\(n=5\)), \emph{build and CI/CD workflow} (\(n=3\)), and \emph{refactoring and migration} (\(n=3\)) remained visible. Overall, the observed plan files supported both implementation work and surrounding engineering activities, but their relative emphasis was shaped by the dominant repository.


Table~\ref{tab:rq2-coding-examples} illustrates how concrete tasks in plan files were mapped to task groups and SWEBOK categories.

\begin{table}[H]
\centering
\caption{Examples of coded plan file purposes from the most frequent SWEBOK areas.}

\label{tab:rq2-coding-examples}
\setlength{\tabcolsep}{2.0pt}
\renewcommand{\arraystretch}{1.02}

\begin{tabularx}{\linewidth}{
>{\RaggedRight\arraybackslash}p{2.4cm}
>{\RaggedRight\arraybackslash}p{2.9cm}
>{\RaggedRight\arraybackslash}p{2.9cm}
>{\RaggedRight\arraybackslash}p{2.6cm}
>{\RaggedRight\arraybackslash}X
}
\toprule
\textbf{Repository} & \textbf{Plan file name} & \textbf{Task-level} & \textbf{Task-group level} & \textbf{SWEBOK} \\
\midrule

\repo{forcedotcom/\\salesforcedx-\\vscode} & \textit{Enable Naming Convention ESLint Rules} & ESLint enforcement of naming conventions. & Code quality & Software Quality \\

\repo{boldgrid/\\w3-total-\\cache} & \textit{travis-php85-gha-migration} & Migration of PHP and WordPress test automation from Travis to GitHub Actions and Argo. & Build and CI/CD workflow & Software Configuration Management \\

\repo{forcedotcom/\\sfdx-core} & \textit{Update Features from Salesforce Docs} & Synchronization of feature metadata from Salesforce documentation. & Developer tooling & Software Engineering Process \\

\repo{adobe/\\spectrum-\\design-data} & \textit{Size Property Analysis Across Component Schemas} & Analysis of size-option patterns across component schemas. & Design system standardization & Software Design \\
\bottomrule
\end{tabularx}

\end{table}


\subsection{RQ3: What Information do Agent Plans Provide?}
\label{sec:results_rq3}

Using the heading and section-content analysis described in Section~\ref{sec:data_analysis}, we examined the information categories that Agent Plans provide to guide the tools. Table~\ref{tab:plan-information-categories} summarizes the resulting information categories. 

\begin{table}[H]
\caption{Information categories in Agent Plans by corpus and AI co-authorship signal. 
  \textbf{All repos}: full corpus (\(n=85\) plan files, 10 repos).
  \textbf{Excl.\ SF}: excluding \texttt{forcedotcom/salesforcedx-vscode}
  (\(n=19\) plan files, 9 repos).
  Occ. = heading occurrences; Files = distinct plan files; Repos = distinct repositories.%
}
\label{tab:plan-information-categories}
\scriptsize
\setlength{\tabcolsep}{3pt}
\begin{tabularx}{\linewidth}{p{0.35\linewidth} r r r r r r r}
\toprule
 & \multicolumn{3}{c}{\textbf{All repos}} & \multicolumn{2}{c}{\textbf{Excl.\ SF repo}} & \multicolumn{2}{c}{\textbf{AI co-authorship}} \\
\cmidrule(lr){2-4}\cmidrule(lr){5-6}\cmidrule(lr){7-8}
\textbf{Category} & \textbf{Occ.} & \textbf{Files} & \textbf{Repos} & \textbf{Occ.} & \textbf{Repos} & \textbf{AI} & \textbf{No signal} \\

\midrule
Implementation steps   & 82 & 54 & 8 & 20 & 7 & 19 & 63 \\[2pt]
Files and locations    & 68 & 46 & 7 & 11 & 6 & 21 & 47 \\[2pt]
Testing and validation & 57 & 40 & 6 & 9  & 5 & 22 & 35 \\[2pt]
Architecture and design & 39 & 26 & 6 & 11 & 5 & 24 & 19 \\[2pt]
Analysis and findings  & 30 & 18 & 4 & 8  & 3 & 14 & 16 \\[2pt]
Project context        & 29 & 29 & 6 & 7  & 5 & 13 & 16 \\[2pt]
Task breakdown         & 28 & 13 & 4 & 5  & 3 & 7  & 19 \\[2pt]
Tooling and build configuration & 24 & 20 & 5 & 10 & 4 & 8 & 16 \\[2pt]
Design constraints and decisions & 23 & 21 & 7 & 7 & 6 & 8 & 15 \\
\bottomrule
\end{tabularx}
\end{table}

Sections that mainly stated the goal, problem, topic, or scope of the plan file were coded as \textit{Plan overview and scope}. We retained this label in the coding output for traceability, but excluded it from Table~\ref{tab:plan-information-categories} because it describes what the plan file is about, rather than how the agent is guided to carry it out. \textit{Implementation steps} sections typically describe actions such as updating files, removing obsolete code, migrating behavior, or implementing planned changes~\cite{youtrackdbPlan}.
\textit{Files and locations} sections narrow the technical scope by identifying packages, modules, tests, schemas, traces, or other artifacts to inspect or modify. \textit{Testing and validation} sections describe checks, tests, success criteria, or expected behavior used to determine whether the task is complete~\cite{botonicPlan}. The remaining categories provide supporting guidance for execution. \textit{Architecture and design} captures architectural context, data flow, and design choices~\cite{spectrumPlan}. \textit{Analysis and findings} records observations, root causes, or patterns from failures and previous runs~\cite{odbDesignPlan}. \textit{Project context} describes the current system state, dependencies, workflow, or environment. \textit{Task breakdown}, \textit{Tooling and build configuration}, and \textit{Design constraints and decisions} organize the work, identify tools and workflows to use, and define boundaries the agent should preserve~\cite{w3cacheMigrationPlan}.

The most common category was \textit{Implementation steps}, followed by \textit{Files and locations} and \textit{Testing and validation}. These categories show that plan files commonly specify what actions to perform, where to perform them, and how to check whether the work is complete. To account for the concentration of plan files in \texttt{forcedotcom/salesforcedx-vscode}, the table also reports heading occurrences and repository coverage, excluding that repository. The same broad information categories remained visible after excluding the dominant repository, although their relative ordering changed. \textit{Implementation steps} remained the most frequent category, while \textit{Files and locations}, \textit{Architecture and design}, and \textit{Tooling and build configuration} were also common in the remaining repositories. This suggests that the broad structure of Agent Plans was not limited to \texttt{salesforcedx-vscode}, but the exact ranking of categories should be interpreted with caution, as only 19 plan files remained after excluding the dominant repository.

We also compared information categories by explicit AI co-authorship signal. This split separates plan files associated with commits that contain explicit AI co-authorship signals from those with no such signal. The split showed that AI-co-authored plan files and plan files without a clear signal used the same broad kinds of execution guidance. The file-level counts show that the main categories were not only produced by a few highly structured files. \textit{Implementation steps}, \textit{Files and locations}, and \textit{Testing and validation} appeared in 54, 46, and 40 distinct plan files, respectively. In the heading-occurrence split by explicit AI co-authorship signal, plan files associated with AI-co-authored commits most often contained headings related to architecture and design, testing and validation, files and locations, and implementation steps. Plan files without explicit AI co-authorship signals often contained headings related to implementation steps, files and locations, and testing and validation. The main plan file structure was similar across the two groups, although the relative emphasis differed.

Overall, Agent Plans translate the planned task into actionable information about changes, locations, validation, context, tools, and constraints. This distinguishes them from repository-level context files, which typically provide standing project-level guidance. Comparing the nine Agent Plan categories with the \texttt{AGENTS.md} content categories reported by Mohsenimofidi et al.~\cite{mohsenimofidi2025context} further illustrates this distinction (Table~\ref{tab:categories-comparison}). Three categories, namely \textit{Implementation steps}, \textit{Files and locations}, and \textit{Task breakdown}, have no direct counterpart in \texttt{AGENTS.md}. The remaining categories partially overlap with the \texttt{AGENTS.md} categories but differ in scope, as Agent Plans encode task-specific knowledge rather than general repository guidance.

\begin{table}[H]
\centering
\caption{Comparison of Agent Plan information categories against
\texttt{AGENTS.md} categories~\cite{mohsenimofidi2025context}.}
\label{tab:categories-comparison}
\begin{tabular}{lll}
\toprule
Agent Plan Category & Closest \texttt{AGENTS.md} Category & Relationship \\
\midrule
Implementation steps             & None                               & Plan-specific \\
Files and locations               & None                               & Plan-specific \\
Task breakdown                    & None                               & Plan-specific \\
\midrule
Analysis and findings             & Troubleshooting                    & Scope differs \\
Design constraints and decisions  & Conventions                        & Scope differs \\
Architecture and design           & Architecture/structure             & Scope differs \\
Testing and validation            & Test execution, Test strategy      & Scope differs \\
Tooling and build configuration   & Build commands, References         & Scope differs \\
Project context                   & Goals/purposes, Setup, Tech stack  & Scope differs \\
\bottomrule
\end{tabular}
\end{table}



\section{Discussion}
\label{sec:discussion}

\subsection{Implications for Research and Future Work}


Agent Plans make intended work explicit so that it can be revised and preserved alongside code. Although public preservation is uncommon, the practice is actively discussed in developer communities~\cite{cantor2025cursor,reddit2025cursorplan}, suggesting that local, uncommitted use is likely more widespread than repository traces indicate. Below we list several implications of our findings.

\begin{itemize}
\item \textbf{For tool builders:} The three most frequent categories: \textit{Implementation steps}, \textit{Files and locations}, and \textit{Testing and validation}, form a minimum viable plan file schema covering what the agent should do, where, and how to verify completion; tools should scaffold these three sections by default. We did not observe any files in \texttt{.gemini/plans/} across the screened repositories, suggesting that future work should examine whether Gemini-related planning artifacts are stored elsewhere or remain local.

\item \textbf{For practitioners:} A plan file lacking any of these three categories is likely under-specified as operational guidance, giving teams a lightweight readiness check; explicit conventions for naming, structuring, and committing plan files are still needed.

\item \textbf{For researchers:} The mining pipeline~\cite{baltes2026pipeline} can be extended to broader directory patterns (e.g., \texttt{docs/plans/}) and complemented by surveys to capture local plan files. The most pressing open question is whether plan files affect outcomes: comparing PR merge time and revert rate for \texttt{forcedotcom/\allowbreak salesforcedx-vscode} commits with and without an associated plan file, controlling for author identity, file and line count, would provide direct evidence of whether planning artifacts improve work. A key open question is whether planning workflows could be extended to cover requirements clarification and design exploration, activities not observed in this corpus, enabling agents to participate earlier in the software lifecycle. Finally, going beyond the \emph{emerging results} in this paper, researchers could explore how developer activities in plan files (RQ2) are supported by the information in plan files (RQ3).
\end{itemize}




\subsection{Threats to Validity}
\begin{itemize}

\item \textbf{Construct validity.} The analysis is restricted to repository-preserved Markdown files found in tool-specific plan directories; plan files that remained local, transient, or stored under other naming schemes are not captured. The directory-pattern search may introduce selection bias. By searching only tool-specific plan directories, we likely missed plan-like artifacts stored under other paths or retained only in local tool state, issues, pull requests, or external planning documents. The reported prevalence should therefore be interpreted as prevalence under the searched repository-directory conventions. Authorship attribution is also uncertain: of the 72 unique plan-file commits for which we retrieved complete metadata, 31 (43.1\%) showed explicit AI co-authorship signals, affecting 27 of the 85 plan files. The remaining plan files should not be interpreted as definitively human-authored, as the absence of an explicit co-authorship signal does not constitute evidence of human authorship.

\item \textbf{Internal validity.} The coding for RQ2 and RQ3 relied on manual interpretation, and alternative category assignments remain possible, particularly for overlapping SWEBOK areas and section-level information categories. Two researchers independently coded all plan files and resolved disagreements through discussion before producing the final coding scheme. Some plan filenames were also insufficiently descriptive, complicating traceability during analysis.

\item \textbf{External validity.} The corpus is small (85 files from 10 repositories) and heavily concentrated: 66 plan files (77.6\%) originate from a single repository, and approximately 78.8\% were contributed by one developer across two related Salesforce repositories, making corpus concentration a stronger threat than small size alone. Sensitivity analyses show that several activity and information categories also appear outside the dominant repository, but generalization beyond the Salesforce ecosystem is limited, and plan files stored under other directory conventions are not represented. A further concern is survivor bias: plan files deleted, added to \texttt{.gitignore}, squashed, or rebased before the scan are invisible to this study.
\end{itemize}


\section{Conclusion}
\label{sec:conclusion}


This \emph{Emerging Results} paper identifies 85 Agent Plan files from 10 repositories across 36{,}710 screened GitHub projects, showing that public preservation is uncommon and heavily concentrated (77.6\% from one repository), while sensitivity analyses show that several activity and information categories also appear outside the dominant repository. Plan files supported a range of activities and exhibited recurring information structures, most prominently implementation steps, files and locations, and testing and validation. Most plan files appeared in only a single commit, which means that they function as task-specific records rather than documents that are repeatedly maintained over time. These findings position repository-preserved Agent Plan files under tool-specific directories as a narrow but informative repository artifact that surfaces a task-oriented layer of human-agent coordination beyond coarse-grained context files, and point to directions for different audiences as agentic AI coding tool adoption continues to grow.

\section{Data Availability}
The repository-processing pipeline used to derive the set of screened repositories, along with the study data and analysis artifacts is available on Zenodo~\cite{baltes2026pipeline, abubakar2026pipeline}.



\bibliography{literature}

\appendix

\end{document}